\documentclass[10pt]{article}
\def\EllipticF{\mathop{\mit F}\nolimits}
\def\EllipticE{\mathop{\mit E}}
\def\cn{\mathop{\rm cn}\nolimits}
\def\sn{\mathop{\rm sn}\nolimits}
\def\csch{\mathop{\rm csch}\nolimits}
\def\EllipticPi{\mathop{\mit \Pi}\nolimits}

\def \bfx{{\bf x}}
\def\veck{\bf k}
\def\veckappa{\mbox{$\kappa\hskip-.6em\kappa$}}

\def \kappax{{\veckappa \cdot \bfx}}
\def\fh{{\frac{1}{2}}}
\def\fq{\frac14}
\def\fsh{{\frac{1}{\sqrt2}}}
\def\fraceb{\frac{\eta}{\beta}}
\def\G14{{\mit\Gamma}(\frac14)}
\def\Faza{\mit\Theta}
\def\Lamit{{\mit\Lambda}}
\def\Omit{{\mit\Omega}}
\def\calM{{\cal M}}

\def\calAveck{{{\cal A}_{\veck}}}
\def\calML{\calM/\Lamit}
\def\uk{{\sf u}_{\veck}}
\def\ukappa{{\sf u}_\kappa}
\def\uveckappa{{\sf u}_{\veckappa}}
\def\omegab{\omega_{\rm b}}
\def\JacobiAmplitude{\mathop{\rm am}\nolimits}
\def\InverseJacobiSN{\mathop{\it F}\nolimits}

\def \Journal#1#2#3#4{{#1} {\bf #2}, #3 (#4)}

\def \ApJ{Astrophys. J.}
\def \JETP{Sov. Phys. JETP}
\def \MNRAS{Mon. Not. R. Astron. Soc.}
\def \ASS{Astroph. Space Sci.}
\def \CQG{Class. Quantum Grav.}
\def \JMP{J. Math. Phys.}
\def \PRD{{Phys. Rev.} D}
\def \Science{Science}

\def\grant{No 2 P03D 014 17}
\begin{document}
\title{Anomalous dispersion of density waves in the early universe
with positive cosmological constant}

\author{Zdzis{\l}aw A. Golda and Andrzej Woszczyna\\
Astronomical Observatory, Jagellonian University\\
ul. Orla 171, 30--244 Krak\'ow, Poland}

\maketitle
\begin{abstract}
Density perturbations in the flat ($K=0$) Robertson-Walker universe with
radiation ($p=\varepsilon/3$) and positive cosmological constant 
($\Lambda>0$) are investigated. The phenomenon of anomalous dispersion of
acoustic waves on $\Lambda$ is discussed.
\end{abstract}

\newpage

\section{Introduction}
The image the perturbations create on the last scattering surface depends on how
they propagate in the radiational era. In the flat radiation dominated
universe with vanishing cosmological constant the density perturbations
form sound waves and propagate with the constant sound velocity $v=1/\sqrt{3}$. They
obey linear dispersion relation $\omega=k/\sqrt{3}$, therefore, their phase
and group velocities are identical. Their propagation does not depend on the wave
number, therefore, no critical scale appear and no gravitationally bound structures form.
The above is the subject of theorem proved by Sachs and Wolfe 
(\cite{Sachs&Wolfe} sec.~II), confirmed independently by Lukash \cite{Lukash}, Chibisov
and Mukhanov \cite{Chibisov&Mukhanov} on the basis of Hamilton formalism in
Field-Shepley variables \cite{Field&Shepley}. The result holds in the gauge-invariant 
descriptions \cite{Golda&Woszczyna2} as well as in the original
Lifshitz formalism \cite{Golda&Woszczyna1}.

The situation changes significantly in the universe with the negative space
curvature ($K=-1$), where the acoustic wave propagates as a scalar field with
the mass $m=-K$.\footnote{Gauge-invariant perturbation equation takes the form
of eq.~(5.4) of \cite{Birrell&Davies} with the minimal coupling ($\xi=0$). The
same is also the form of eq.~(21) of \cite{Sachs&Wolfe} and eq.~(4.6) of \cite{
Chibisov&Mukhanov}.} The dispersion relation is nonlinear and, in consequence,
some critical frequency $\omega_{\rm cr}$ appear. Below $\omega_c$ the wave
propagation is forbidden. The dispersion of the acoustic wave on the curvature
affects the microwave background fluctuations and in principle can be used
to measure the cosmological density parameter $\Omit$ \cite{Golda&Woszczyna2}.

In this letter we show that also the cosmological constant $\Lamit$ is a
source of dispersive phenomena. The interest in $\Lamit$-cosmology grows, owing to
speculations that cosmological constant may take substantial values in the 
present epoch~\cite{Bahcall&Ostriker&Perlmutter&Steinhardt}. We investigate the
flat ($K=0$) universe with radiation ($p=\varepsilon/3$) and positive
cosmological constant ($\Lamit>0$). We find exact solutions for gauge-invariant 
perturbation equations \cite{Bardeen}--%
\cite{Jackson}
and discuss their properties to show, that in the presence of positive
$\Lamit$ the dispersion of the acoustic field has an anomalous character.

\section{Homogeneous background}

In the flat radiation-dominated universe with a positive cosmological
constant, Friedman equations are satisfied by the scale factor $a(t)$:
	\begin{equation}
a(t) = \left(\calML\right)^{1/4}
\sqrt{\sinh(2\sqrt{\Lamit/3}\,t)} = \left(\calML\right) ^{1/4}
\sqrt{\sinh \tau}
	\label{eq1}
	\end{equation}
where $\calM$ is the constant of motion ${\calM}=\epsilon\, a^4$, $\epsilon$
means the energy density and the dimensionless time parameter~$\tau$ is
related to the metric time as $\tau=2 \sqrt{\Lamit/3}\,t$. To simplify
notation in the forthcoming formulae we also introduce the constant $\beta$
defined as $\beta = \sqrt{3} /[16\calM\Lamit]^{1/4}$. The conformal time
$\eta$
	\begin{equation}
\eta =\EllipticF\left(\arccos\left[\frac{1-\sinh
\tau}{1+\sinh\tau}\right],\frac{1}{\sqrt{2}}\right) \beta 
	\label{eq2}
	\end{equation}
is finite
$\eta\in [0, \, 2K({ 2^{-1/2}})\beta] = [0, \,\fh\pi^{-1/2}[\Gamma(\fq)]^2\beta]$, 
where $\EllipticF(\varphi,m)$ stands for the elliptic integral of the first
kind and $K(m)$ is the complete elliptic integral of the first kind 
\cite{Byrd&Friedman}. The scale factor $a(\eta)$ expresses by the Jacobi elliptic
functions $\sn(u, m)$, $\cn(u, m)$ of the ratio $\eta/\beta$.
	\begin{equation}
a(\eta) = 2 \beta\sqrt{\calM/3}\,\frac{\sn(\fraceb,\fsh)}{1 + \cn(\fraceb,\fsh)}.
	\label{eq3}
	\end{equation}
The scale factor $a(\eta)$ grows monotonically from $0$ to infinity. In the
neighborhood of the initial singularity the radiation is dynamically dominant
and the effect of the cosmological constant can be omitted. The opposite
boundary of the conformal time interval 
($\eta =\fh\pi^{-1/2}[\Gamma(\fq)]^2\beta$) correspond to the de Sitter epoch
where $\Lamit$ dominates and the radiation plays a marginal role.

\section{The perturbation equation}

By variating Raychaudhuri and the continuity equations one obtains the
propagation equation for the density contrast $\delta\epsilon/\epsilon$. While
$\delta\epsilon/\epsilon$ is measured on the flow-orthogonal hypersurfaces 
\cite{Lyth&Mukherjee} the equation reads
	\begin{equation}
-\fh\coth^2(\tau) X(\tau, \bfx) + \fh \coth(\tau)
X'(\tau, \bfx) + X''(\tau) = v^2\beta^2\csch(\tau) \nabla^2X(\tau, \bfx).
	\label{eq4}
	\end{equation}
$\nabla^2$ denotes the Laplace operator in the conformal space,
consequently its eigenvalue $-k^2$ is related to the time invariant ``commoving'' 
wave vector ${\veck}$ with $k=|{\veck}|$. The same equation can be obtained in
other gauge-invariant theories \cite{Bardeen}--%
\cite{Olson}, 
\cite{Woszczyna&Kulak}--%
\cite{Jackson} under necessary
redefinition of the perturbation variables\footnote{Although we restrict
ourselves to $v=1/\sqrt3$ we explicitly keep $v$ in perturbation equations to
make clear their structure and correspondence to other formalisms.} (see 
\cite{Golda&Woszczyna2}).
This is convenient to express the equation (\ref{eq4}) in terms of the
conformal time 
	\begin{equation}
\frac{1}{\beta^2} 
\left(
1-\frac{2}{\sn^2(\fraceb,\fsh)}
\right) 
X(\eta, \bfx) - v^2\,\nabla^2 X(\eta, \bfx)
+ X''(\eta, \bfx)=0 
	\label{eq5}
	\end{equation} 
and search for its solutions in the form of the Fourier integral
	\begin{equation}
X(\eta, \bfx) = \int\calAveck\, \uk(\eta, \bfx)\,d^3k + \mbox{c.c.}
	\label{eq6}
	\end{equation}
Prime stands for the conformal time derivative, 
Fourier coefficients ${\calAveck}$ are arbitrary functions of the wave number $k$.
Modes $\uk(\eta, \bfx)$ satisfy both, the
Helmholtz equation
	\begin{equation}
\nabla^2 \uk(\eta, \bfx) = -k^2\uk(\eta, \bfx)
	\label{eq7}
	\end{equation}
and the time equation (\ref{eq5}), and divide into two classes 
distinguished by the critical value
$k=\frac{1}{\sqrt{2/3}\,\beta}=(4{\calM}\Lamit)^{1/4}$.
Below, we employ the
dimensionless wave vector $|{\veckappa}|=\kappa=\sqrt{2}\, v \beta k=
\sqrt{2/3}\,\beta k$. In this notation both classes of solutions
are defined by inequalities $\kappa>1$ and $\kappa\leq1$. They will be 
discussed separately.

\section{Subcritical perturbations: solutions for $\kappa > 1$}

The short-scale solutions ($\kappa>1$) propagate as waves of 
variable amplitude and variable frequency. Elementary solutions 
$\uveckappa(\eta,\bfx)$ take the form 
	\begin{equation}
\uveckappa(\eta,\bfx)
=\frac{\sn([2\sqrt{\pi}\,\beta]^{-1}[\G14]^2,\fsh) \sqrt{2 +
(\kappa^2-1)\sn^2(\fraceb,\fsh)}} 
{\sqrt{2+(\kappa^2-1)\sn^2([2\sqrt{\pi}\,\beta]^{-1}
[\G14]^2,\fsh)}\,\sn^2(\fraceb,\fsh)}
\exp[i\,\Faza(\eta,\bfx)]
	\label{eq8}
	\end{equation}
where the phase $\Faza(\eta,\bfx)$ is given by
	\begin{equation}
\Faza(\eta,\bfx) = \frac{\kappax}{\sqrt{2}\beta v} - 
\frac{\kappa\sqrt{\kappa^4-1}}{\sqrt2(\kappa^2-1)} 
\left[\fraceb-\EllipticPi\left(
\JacobiAmplitude\left(\fraceb,\fsh\right),-\frac12(\kappa^2-1),\fsh\right)
\right].
	\label {eq9}
	\end{equation}
Function $\EllipticPi(\varphi,\alpha^2,m)$ is the elliptic integral of the third
kind and $\JacobiAmplitude(u,m)$ is the amplitude elliptic
function~ \cite{Byrd&Friedman}. 
The solution is singular at both ends of the
conformal time interval. In the vicinity of the initial singularity
($\eta=0$) the radiation plays the dominant role and the solutions
$\uk(\eta,\bfx)$ approach appropriate modes for the universe with
vanishing $\Lamit$ \cite{Golda&Woszczyna2}. In the late time regime
$\eta\simeq\frac{\beta}{2\sqrt\pi}[\G14]^2$ the universe
realizes the exponential growth under the dominant effect of
$\Lamit$. Singularities in solutions $\uk(\eta, \bfx)$ in the
$\eta\to\frac{\beta}{2\sqrt\pi}[\G14]^2$ limit is known as
the instability of de Sitter universe \cite{Zimdahl}.

The local frequency\footnote{see \cite{Whitham}}
$\omega(\eta,\kappa)$, evaluated as the time derivative of 
the phase $\Faza(\eta,\bfx)$  
	\begin{equation}
\omega(\eta,\kappa)=\frac{\partial\,\Faza(\eta,\bfx)}{\partial\,\eta}
= \frac{\kappa\sqrt{\kappa^4-1}\sn^2(\fraceb,\fsh)}
{\sqrt2\beta(2+(\kappa^2-1)\sn^2(\fraceb,\fsh))}.
	\label{eq10}
	\end{equation}
is a time-dependent quantity. The interval $\eta\in 
[\eta_1,\,\,\eta_2]$ contained between two extrema of the second derivative
$\frac{\partial^2\omega(\eta,\kappa)}{\partial\eta^2}$
\[
\eta_1=\InverseJacobiSN
\left(
\arcsin
\frac{\sqrt2}{\sqrt{1+\kappa^2}}
,\fsh
\right)\beta,\qquad\eta_2=\fh\pi^{-1/2}[\textstyle\G14]^2\beta-\InverseJacobiSN
\left(
\arcsin
\frac{\sqrt2}{\sqrt{1+\kappa^2}},\fsh
\right)\beta
\]
form the oscillatory stage, where the perturbation propagates as the
acoustic wave. $\omega(\eta,\kappa)$ maintains an approximately constant value
there. The quantity 
	\begin{equation}
\omegab(\kappa) = \frac{\kappa\sqrt{\kappa^4-1}}{\sqrt2\beta(\kappa^2+1)}.
	\label{eq11}
	\end{equation}
may be understood as the basic frequency, while
its the modulation $\omega_{\rm mod}(\eta,\kappa)$ is given by
	\begin{equation}
\omega_{\rm mod}(\eta, \kappa)
 = \frac{\sqrt2\kappa\sqrt{ \kappa^4-1}(-1 + \sn^2(\fraceb, \
\fsh))}{\beta(1 + \kappa^2)(2 + (\kappa^2-1)\sn^2(\fraceb, \fsh))}.
	\label{eq12}
	\end{equation}
Beyond this interval $[\eta_1,\,\,\eta_2]$ the division
into basic frequency and the modulation loses its physical 
meaning. While $\eta\to0$ or
$\eta\to\frac{\beta}{2\sqrt\pi}[\G14]^2$ the frequency formally
defined in (\ref{eq10}) tends to zero  and the
amplitude grows indefinitely. In fact, the
concept of oscillation, breaks down in these limits.

Formula (\ref{eq10}) plays the role of time-dependent dispersion relation, and
(\ref{eq11}) is its analog for the basic frequency. Both relations are nonlinear.
The number $\kappa=1$ ($k = \sqrt{2/3}\,\beta = [4{\calM}\Lamit]^{1/ 4}$) forms 
the critical value of the wave number, and naturally defines the critical
wave-length $\lambda_{\rm cr}=\frac{\sqrt2\pi}{[{\calM}\Lamit]^{1/4}}$. 
Below this scale the wave propagation is forbidden. While $\kappa\to1$, the 
frequency tends to zero, therefore, perturbations longer than the critical
size do not form travelling waves. In contrast to dispersion on the space
curvature, the dispersion on $\Lamit$ manifests its anomalous character: critical 
behaviour concerns the scale of the perturbation (the wave number $k$), not the 
frequency\footnote{The space curvature defines the minimal frequency
like the plasma frequency.} $\omega$. The anomaly is also clear from 
the phase and the group velocities behaviour
	\begin{equation}
v_{\rm f}(\eta, \kappa, \beta) = \frac{\omega(\eta,\kappa)}{k} = 
\frac{\sqrt{\kappa^4-1}\sn^2(\fraceb,\fsh)}{\sqrt3\,[2 + (\kappa^2-1) 
\sn^2(\fraceb, \fsh)]},
	\label{eq13}
	\end{equation}
	\begin{equation}
v_{\rm g}(\eta,\kappa,\beta) = \frac{\partial\,\omega(\eta,\kappa)}{\partial\,k} = 
\frac{[6\kappa^4-2+(\kappa^2-1)(\kappa^4-2\kappa^2-1)\sn^2(\fraceb,\fsh)]
\sn^2(\fraceb,\fsh)}
{\sqrt3\,\sqrt{\kappa^4-1}\,[2+(\kappa^2-1)\sn^2(\fraceb,\fsh)]^2}.
	\label{eq14}
	\end{equation}
The phase velocity $v_{\rm f}(\eta, \kappa, \beta)$ decreases with increasing wave-length
\cite{Szczeniowski} and, in consequence, the group velocity $v_{\rm
g}(\eta, \kappa, \beta)$  is always greater than
the phase velocity $v_{\rm f}(\eta, \kappa, \beta)$. 
In the $\kappa\to1$ limit the phase velocity
tends to zero, while the group velocity formally expressed as
(\ref{eq14}) grows indefinitely. In this regime the concept of
the wave packet loses its sense.\footnote{Taylor series for $\omega$
cannot be cut at linear terms. Higher order terms are
responsible for the diffusion of the wave packets \cite{Ginzburg}.}

\section{Supercritical perturbations: solutions for $\kappa\leq1$}

The large-scale solutions ($\kappa\leq1$) do not propagate as
travelling waves. The space of solutions consists of the
combinations of the two modes $\ukappa^{(1)} \exp(i\, \kappax)$ and
$\ukappa^{(2)}\exp(i\, \kappax)$, where both $\ukappa^{(1)}$ and $\ukappa^{(2)}$
	\begin{eqnarray}
\ukappa^{(1)}(\eta)	&=&
		\frac{\sn([2\sqrt{\pi}\,\beta]^{-1}[\G14]^2,\fsh)\, 
\sqrt{2 - (1-\kappa^2)\sn^2(\fraceb,\fsh)}}{\sqrt{2-(1-\kappa^2)\sn^2([2\sqrt{\pi}\,\beta]^{-1}[\G14]^2,\fsh)}\,\sn(\fraceb,\fsh)}\nonumber\\
			&&\exp
\left[
\frac{\kappa\sqrt{1-\kappa^4}}{\sqrt2(1-\kappa^2)} 
\left(-
\fraceb +
\EllipticPi
\left(
\JacobiAmplitude\left(\fraceb,\fsh
\right),
\fh\left(1-\kappa^2\right), \fsh\right)
\right)
\right],\\[10pt]
	\label{eq15}
\ukappa^{(2)}(\eta)	&=&\ukappa^{(1)}(\eta)\int\limits_0^\eta
\frac{1}{[\ukappa^{(1)}(\eta')]^2}d\eta'\, .
	\label{eq16}
	\end{eqnarray} 
are real functions. In the vicinity of the initial singularity $\ukappa^{(1)}$ behave as
the decaying mode, while $\ukappa^{(2)}$ plays the role of the
growing one. In this particular sense the behaviour of the
large-scale perturbations mimics that of modes in the
pressureless matter. Later on the solutions evolve differently.
Both modes keep approximately constant amplitude in the
epoch when the $\kappa>1$ modes oscillate. Finally, in the de
Sitter regime both modes $\ukappa^{(1)}$ and $\ukappa^{(2)}$ blow up
exponentially. 

The two limit cases $\kappa=0$ and $\kappa=1$ express in simple formulae
	\begin{eqnarray}
\ukappa^{(1)}(\eta) &=&
\frac{\sn([2\sqrt{\pi}\,\beta]^{-1}[\G14]^2,\fsh)} 
{\sqrt{2-\sn^2([2\sqrt{\pi}\,\beta]^{-1}[\G14]^2,\fsh)}}
\frac{\sqrt{2-\sn^2\left(\fraceb,\fsh\right)}}{\sn\left(\fraceb,\fsh\right)},
	\label{eq17}\\[1ex]
\ukappa^{(2)}(\eta) &=&\frac{\sn([2\sqrt{\pi}\,\beta]^{-1}[\G14]^2,\fsh)} 
{\sqrt{2-\sn^2([2\sqrt{\pi}\,\beta]^{-1}[\G14]^2,\fsh)}}\nonumber\\
&&\left[
\frac{\sqrt{2-\sn^2\left(\fraceb,\fsh\right)}}{\sqrt2\sn\left(\fraceb,\fsh\right)}
\left[2 \EllipticE \left(\JacobiAmplitude\left(\fraceb,\fsh\right),\fsh\right)-\fraceb\right]-\cn\left(\fraceb,\fsh\right)
\right]
	\label{eq18}
	\end{eqnarray} 
for $\kappa=0$ and
	\begin{eqnarray}
\ukappa^{(1)}(\eta) &=&\sn\left([2\sqrt{\pi}\,\beta]^{-1}[{\textstyle\G14}]^2,\fsh\right)\frac{1}{\sn(\fraceb,\fsh)},
	\label{eq19}\\
\ukappa^{(2)}(\eta) &=&\sn\left([2\sqrt{\pi}\,\beta]^{-1}[{\textstyle\G14}]^2,\fsh\right)
\frac{1}{\sn(\fraceb,\fsh)}
\left[
\fraceb -\EllipticE \left(\JacobiAmplitude\left(\fraceb,\fsh\right),\fsh\right)
\right]
	\label{eq20}
	\end{eqnarray}
for $\kappa=1$.

The dynamics of large-scale perturbations does not confirm their
exceptional role in the structure formation process: their amplitudes
are of the same range as those of oscillating modes. However,
since these perturbations do not propagate, the density and expansion
excesses are correlated to each other throughout the space  for
decaying modes, and anti- correlated for the growing ones. Present
observations of the microwave background, which are limited to the
single physical quantity (temperature), are not complete enough to
distinguish between traveling waves and the perturbations of other
types. Yet it is possible that we will get more complete
data from the last scattering surface in future. Therefore, some
estimations of linear diameters and appropriate angular scales on the
sky are worth mentioning.

We assume that the universe after decoupling evolves as filled
with presureless matter. Then the lapse of conformal time
between the emission and the observation moments is given by 
	\begin{equation}
\Delta\eta(z) = 
\sqrt{3/\Lamit} \left[\frac{\epsilon_0}{m}\right]^{1/3}
\left((1+z)\, 
{}_2F_1\left[\frac13,\fh;\frac43;-(1+z)^3\frac{\epsilon_0}{\Lamit}\right]-
{}_2F_1\left[\frac13,\fh;\frac43;\frac{\epsilon_0}{\Lamit}\right]\right)
	\label{eq21}
	\end{equation}
where $z$ means the redshift of the last scattering surface,
$m = \epsilon\, a^3$ is the constant of motion for the late
universe, subscript ${}_0$ refers to the present time, and ${}_2F_1[a,b;c;x]$ stands for 
the hypergeometric function \cite{Gradshteyn&Ryzhik}. Sky equator covers $n$
regions of the critical length diameter where
	\begin{equation}
n\left(\frac{\epsilon_0}{\Lamit},z\right) = 
\frac{2\pi\Delta\eta}{\lambda_{\rm cr}}
= \sqrt6\left[
\frac{\epsilon_0}{\Lamit(1 + z)}\right]^{1/4}
\left((1 + z)\,{}_2F_1 \left[\frac13,\fh;\frac43;-(1 +z)^3\frac{\epsilon_0}{\Lamit}\right]
-{}_2F_1\left[\frac13,\fh;\frac43;-\frac{\epsilon_0}{\Lamit}\right]\right).
	\label{eq22}
	\end{equation}
For $\epsilon_0/\Lamit=0.2$ and $z=100$ the number $n$ is about
$n=0.8$, therefore, supercritical perturbations may contribute
to the dipole anisotropy of the MBR. If the radiation plays its
dynamical role after decoupling the number $n$ can be greater,
but always less than $n=2$.

\section{Conclusions}

Like the space curvature, the cosmological
constant causes dispersion of the acoustic field in the early
universe. In both cases the dispersive phenomena are induced by
the space-time geometry and depend neither on the initial
conditions for acoustic field nor on the character of
cosmological epochs prior to the radiational era. Dispersion on
the cosmological constant is a measurable effect in the sense
that appropriate critical length scale is comparable with the
region of the last scattering surface we observe today.
Nevertheless, to detect dispersion one needs more complete
data from the last scattering epoch, in particular, the second
dynamically independent quantity should be known (for instance
the velocity field). In contrast to dispersion on the
space curvature, the dispersion on $\Lamit$ is anomalous. The two
phenomena differ enough to uniquely distinguish their
provenience. Both might be employed to measure values of
$\Lamit$ and $\Omit$ respectively.

\section*{Acknowledgements}

This work was partially supported by State Committee for
Scientific Research, project \grant.


\end{document}